\documentclass[conference]{IEEEtran}
\IEEEoverridecommandlockouts

\usepackage{cite}
\usepackage{amsmath,amssymb,amsfonts}
\usepackage[ruled,vlined,linesnumbered]{algorithm2e}
\usepackage{textcomp}
\usepackage{subcaption}
\usepackage{mwe}
\usepackage{xcolor}
\usepackage[a4paper,
            left=0.75in,
            right=0.75in,
            top=0.773in,
            bottom=1.78in]{geometry}

\usepackage[font=small]{caption}

\def\BibTeX{{\rm B\kern-.05em{\sc i\kern-.025em b}\kern-.08em
    T\kern-.1667em\lower.7ex\hbox{E}\kern-.125emX}}
\begin{document}

\title{\vspace{-1mm} Foundation Model-Aided Hierarchical Deep Reinforcement Learning for Blockage-Aware Link in RIS-Assisted Networks\\ \vspace{-4mm}}

\author{\IEEEauthorblockN{Mohammad Ghassemi\textsuperscript{1}, Han Zhang\textsuperscript{1}, Ali Afana\textsuperscript{2}, Akram Bin Sediq\textsuperscript{2}, \\ and Melike Erol-Kantarci\textsuperscript{1}, \textit{Fellow}, \textit{IEEE}}
\IEEEauthorblockA{\textit{\textsuperscript{1}School of Electrical Engineering and Computer Science, University of Ottawa, Ottawa, Canada} \\
\textit{\textsuperscript{2}Ericsson Inc., Ottawa, Canada}\\
Emails:\{mghas017, hzhan363, melike.erolkantarci\}@uottawa.ca,
\{ali.afana, akram.bin.sediq\}@ericsson.com}

\vspace{-10mm}}
\maketitle

\begin{abstract}
Reconfigurable intelligent surface (RIS) technology has the potential to significantly enhance the spectral efficiency (SE) of 6G wireless networks. However, practical deployment remains constrained by challenges in accurate channel estimation and control optimization under dynamic conditions.
This paper presents a foundation model-aided hierarchical deep reinforcement learning (FM-HDRL) framework designed for joint beamforming and phase-shift optimization in RIS-assisted wireless networks. 
To implement this, we first fine-tune a pre-trained large wireless model (LWM) to translate raw channel data into low-dimensional, context-aware channel state information (CSI) embeddings.
Next, these embeddings are combined with user location information and blockage status to select the optimal communication path. The resulting features are then fed into an HDRL model, assumed to be implemented at a centralized controller, which jointly optimizes the base station (BS) beamforming vectors and the RIS phase-shift configurations to maximize SE.
Simulation results demonstrate that the proposed FM-HDRL framework consistently outperforms baseline methods in terms of convergence speed, spectral efficiency, and scalability.
According to the simulation results, our proposed method improves 7.82\% SE compared to the FM-aided deep reinforcement learning (FM-DRL) approach and a substantial enhancement of about 48.66\% relative to the beam sweeping approach.
\end{abstract}


%%%%%%%%%%%%%%%%%%%%%%%%%%%%%%%%%%%%%%%%%%%%%%%%%%%%%%%%%%%%

\vspace{-1mm}
\section{Introduction}
\vspace{-1mm}
The evolution toward 6G networks has increased reliance on millimeter-wave (mmWave) communication due to its abundant spectral resources. However, mmWave signals are highly susceptible to blockages, which pose significant challenges to communication reliability. Meanwhile, reconfigurable intelligent surfaces (RISs) have emerged as a transformative solution to overcome the limitations of mmWave links by redirecting signals around blockages \cite{direnzo2020smart, wu2019intelligent}. 
RISs effectively enhance wireless coverage and spectral efficiency (SE). 
However, their practical deployment is challenged by the overhead of channel state information (CSI) acquisition and the complexity of BS beamforming and RIS configuration in dynamic environments \cite{pan2020intelligent, wu2020joint, ghassemi2024multi}.

One promising solution to the challenges of CSI acquisition and resource optimization is to integrate foundation models (FMs), such as the large wireless model (LWM) \cite{alikhani2024large}, into RIS-assisted networks. LWM is a pre-trained model on massive wireless channel datasets that can learn fundamental properties of radio propagation. Fine-tuned on data from a specific RIS-assisted scenario, LWM can generate compact, context-aware CSI embeddings that reflect complex interactions such as multipath propagation and BS-RIS-user interaction \cite{ghassemi2025foundation}. These embeddings reduce reliance on pilot-based estimation and support efficient decision-making with minimal signaling overhead \cite{alrabeiah2020deep, pan2025large}.

Deep reinforcement learning (DRL) has also been widely studied for resource optimization in wireless networks due to its ability to handle dynamic environments and complex decision-making processes. However, DRL methods often face significant challenges in high-dimensional settings due to the disparate time scales associated with their inputs \cite{zhu2022drl}. 
To address this, hierarchical deep reinforcement learning (HDRL) decomposes complex control problems in dynamically changing wireless environments into multi-level sub-tasks \cite{zhou2022hierarchical, zhou2024cooperative}.

In this paper, we propose FM-HDRL, an integrated framework that integrates FMs with HDRL. First, we fine-tune the LWM \cite{alikhani2024large} to generate concise CSI embeddings. Next, these embeddings, along with user location and blockage status, are fed into the HDRL agent for joint BS beamforming vectors and RIS phase shift optimization. The main contributions of this work are summarized as follows:
\vspace{-1mm}
\begin{itemize}
    \item We design an HDRL framework where a high-level controller determines the transmission strategy (direct or RIS-assisted) based on spatial blockage information, and a low-level controller optimizes BS and RIS configurations to maximize SE.
    \item We fine-tune a pre-trained open-source FM for RIS-assisted environments to produce compact CSI representations that enhance observability and reduce estimation overhead.
    \item We demonstrate through simulations that the FM-HDRL approach outperforms FM-aided DRL (FM-DRL) and beam sweeping methods in terms of convergence speed, SE, and scalability with increasing the number of RIS elements.
\end{itemize}

The remainder of the paper is organized as follows. Section II reviews related work. Section III presents the system model and problem formulation. Section IV introduces the FM-HDRL algorithm. Section V discusses simulation results. Finally, Section VI concludes the paper.

%%%%%%%%%%%%%%%%%%%%%%%%%%%%%%%%%%%%%%%%%%%%%%%%%%%%%%%%%%%%

\vspace{-1mm}
\section{Related work}
\vspace{-1mm}

Recent studies have explored the potential of FMs in wireless communications. In \cite{alikhani2024large}, an LWM was proposed to generate rich and contextualized channel embeddings by pre-training on large-scale channel datasets. Similarly, BERT4MIMO was developed in \cite{catak2025bert4mimo} to support multiple-input multiple-output (MIMO) systems using transformer-based architectures.
Some existing studies have further adapted generative models for wireless communication tasks such as blockage prediction \cite{ghassemi2025generative}. However, these studies focus on channel representation or environment prediction, without integration into closed-loop control.
Traditional optimization methods have been studied for RIS-assisted networks. For example, \cite{guo2020weighted} proposed a weighted MMSE approach for joint beamforming in RIS-aided systems. Despite their theoretical guarantees, such methods rely on perfect CSI and incur high computational complexity in dynamic environments.
On a parallel track, reinforcement learning techniques have been extensively used to optimize wireless network control. In \cite{zhu2022drl}, a DRL framework was proposed for RIS configuration, demonstrating significant improvements in beamforming efficiency.
HDRL has also been explored to address the challenges of real-time radio access network (RAN) control. The authors in \cite{zhou2024cooperative} introduced a cooperative multi-agent HDRL scheme to enhance energy efficiency in dense networks, while \cite{zhang2025hierarchical} developed a distributed HDRL approach to reduce computational overhead in dynamic user scheduling.
Despite their effectiveness, these methods mainly depend on raw CSI, limiting their scalability in high-dimensional environments due to higher training complexity and sensitivity to imperfect data.

In our prior work \cite{ghassemi2025foundation}, we investigated beam management for mmWave networks by integrating a fine-tuned FM with a DRL agent. Specifically, we demonstrated that the ability of the FM to extract high-level features from raw channel data enabled more efficient beam alignment and reduced training overhead. This paper extends our earlier approach by introducing a hierarchical control framework that uses CSI embeddings from LWM and feeds them into an HDRL agent to jointly optimize BS beamforming and RIS configurations to maximize SE in wireless networks.

%%%%%%%%%%%%%%%%%%%%%%%%%%%%%%%%%%%%%%%%%%%%%%%%%%%%%%%%%%%%

\vspace{-1mm}
\section{System Model and Problem Formulation}
\vspace{-1mm}

In this section, we present the system model for the RIS-assisted wireless network. We begin with an overview of the system architecture, followed by a detailed description of the channel and signal models. Finally, we formulate the problem as an optimization task to maximize SE.

\vspace{-1mm}
\subsection{System Overview}
\vspace{-1mm}

\begin{figure}
    \centering
    \includegraphics[width=0.8\linewidth]{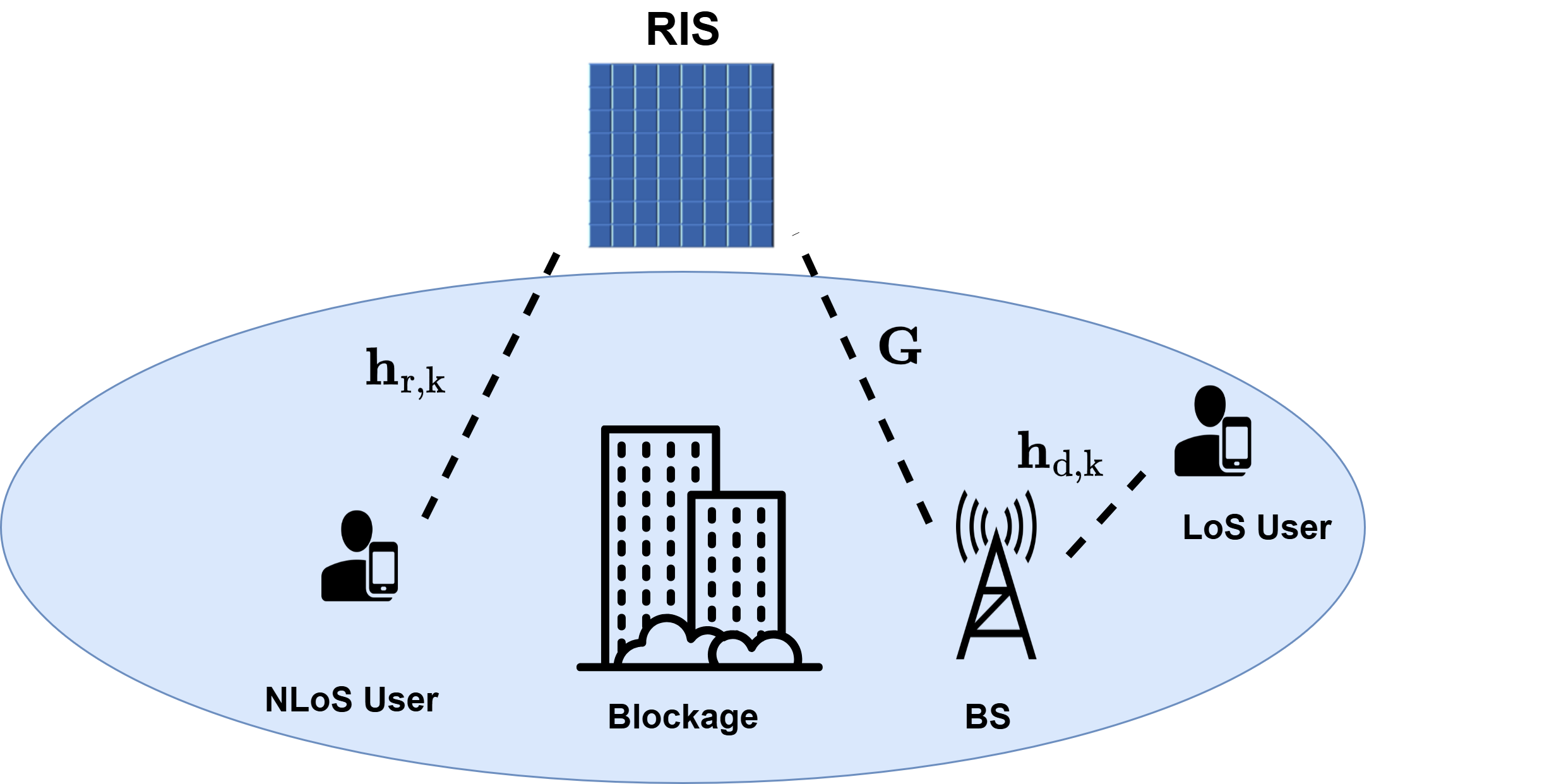}
    \caption{System model of the RIS-assisted communication network.}
    \label{fig:image1}
    \vspace{-4mm}
\end{figure}

In this work, we consider a downlink multi-user communication system as illustrated in \figurename~\ref{fig:image1}. The system comprises a BS equipped with $N$ antennas, serving $K$ single-antenna users. The environment also contains a RIS\textcolor{blue}{,} with $M$ passive reflecting elements. We assume that users are non-stationary and their positions change over time. Consequently, the communication channels are also time-varying. A blockage happens when the communication link between the BS and a user is obstructed by a physical obstacle. If the link is blocked, the RIS provides an alternative communication path. Hence, a direct link is defined as line-of-sight (LoS), while a blocked link corresponds to non-line-of-sight (NLoS).

\vspace{-1mm}
\subsection{Channel and Signal Model}
\vspace{-1mm}
The communication channels in the system consist of three distinct links: the direct BS-to-user link, denoted by $\mathbf{h}_{d,k} \in \mathbb{C}^{N \times 1}$, the BS-to-RIS link, denoted by $\mathbf{G} \in \mathbb{C}^{M \times N}$, and the RIS-to-user link, denoted by $\mathbf{h}_{r,k} \in \mathbb{C}^{M \times 1}$. The RIS can impose controllable phase shifts on the incident signals, which is modeled by the diagonal phase-shift matrix $\mathbf{\Theta} = \text{diag}(e^{j\theta_1}, e^{j\theta_2}, \dots, e^{j\theta_M}) \in \mathbb{C}^{M \times M}$.
A binary indicator $b_k \in \{0, 1\}$ is introduced for each user $k$ to model the link status, where $b_k = 1$ indicates a LoS path and $b_k = 0$ corresponds to an NLoS path.
The effective channel for user $k$ is modeled as \cite{wu2019intelligent}:
\vspace{-1mm}
\begin{equation}
    \mathbf{h}_{\text{eff},k}^H = b_k \mathbf{h}_{d,k}^H + \mathbf{h}_{r,k}^H \mathbf{\Theta} \mathbf{G},\label{eq:formula1}
    \vspace{-1mm}
\end{equation}
where $(\cdot)^H$ denotes the Hermitian transpose.
The BS employs a linear precoding vector $\mathbf{w}_k \in \mathbb{C}^{N \times 1}$ for user $k$. Therefore, the signal received at user $k$ is then given by:
\vspace{-2mm}
\begin{equation}
    y_k = \underbrace{\mathbf{h}_{\text{eff},k}^H \mathbf{w}_k s_k}_{\text{Desired Signal}} + \underbrace{\sum_{j=1, j \neq k}^{K} \mathbf{h}_{\text{eff},k}^H \mathbf{w}_j s_j}_{\text{Multi-user Interference}} + n_k,
    \vspace{-1mm}
\end{equation}
where $s_k$ denotes the data symbol for user $k$, with normalized power $\mathbb{E}[|s_k|^2] = 1$, and $n_k \sim \mathcal{CN}(0, \sigma_k^2)$ is the additive white Gaussian noise at receiver $k$ with noise power $\sigma_k^2$.
Therefore, the signal-to-interference-plus-noise ratio (SINR) experienced by user $k$ is given by:
\vspace{-2mm}
\begin{equation}
    \gamma_k = \frac{|\mathbf{h}_{\text{eff},k}^H \mathbf{w}_k|^2}{\sum_{j=1, j \neq k}^{K} |\mathbf{h}_{\text{eff},k}^H \mathbf{w}_j|^2 + \sigma_k^2}.
    \vspace{-1mm}
\end{equation}

\vspace{-1mm}
\subsection{Problem Formulation}
\vspace{-1mm}
The achievable SE for user $k$ is given by $R_k = \log_2(1 + \gamma_k)$. Our goal is to maximize the sum SE by jointly optimizing the BS beamforming matrix $\mathbf{W} \in \mathbb{C}^{N \times K}$, where each column represents the beamforming vector for a user, and the RIS phase-shift matrix $\mathbf{\Theta}$. The problem is formulated as:
\vspace{-2mm}
\begin{subequations}\label{eq:opt_problem}
\begin{align}
    \text{(P1)}: \max_{\mathbf{W}, \mathbf{\Theta}} \quad & \sum_{k=1}^{K} R_k \label{eq:objective} \\
    \text{s.t.} \quad & \sum_{k=1}^{K} \|\mathbf{w}_k\|^2 \leq P_{\text{max}}, \label{eq:power_constraint} \\
    & |\theta_m| = 1, \quad \forall m \in \{1, \dots, M\}, \label{eq:ris_constraint} \\
    & R_k \geq R_{\text{min}}, \quad \forall k \in \{1, \dots, K\}, \label{eq:fairness_constraint}
\end{align}
\vspace{-1mm}
\end{subequations}
where constraint (\ref{eq:power_constraint}) limits the BS transmit power to $P_{\text{max}}$. Constraint (\ref{eq:ris_constraint}) enforces the unit-modulus condition on RIS elements due to hardware limitations. 
Constraint~(\ref{eq:fairness_constraint}) ensures user fairness by maintaining the SE for each user k above a minimum threshold $R_{\text{min}}$.

\vspace{-1mm}
The optimization problem (P1) is inherently challenging to solve using conventional optimization techniques. The objective function is non-convex because the variables $\mathbf{W}$ and $\mathbf{\Theta}$ are jointly involved in the SINR expression.
Moreover, optimal control policy is fundamentally dependent on blockage indicators ${b_k}$, which evolve slowly compared to fast fading channels. This natural separation of timescales motivates a hierarchical learning approach, where:
\vspace{-1mm}
\begin{itemize}
\item The high-level meta-controller handles slow-timescale blockage events (LoS/NLoS mode selection).
\item The low-level sub-controller performs fast-timescale optimization of $\mathbf{W}$ and $\mathbf{\Theta}$ for the selected mode.
\end{itemize}

This decomposition matches control timescales to their respective channel dynamics.

%%%%%%%%%%%%%%%%%%%%%%%%%%%%%%%%%%%%%%%%%%%%%

\vspace{-1mm}
\section{FM-HDRL for RIS-Assisted Wireless Communication}
\vspace{-1mm}

\begin{figure}
    \centering
    \includegraphics[width=0.8\linewidth]{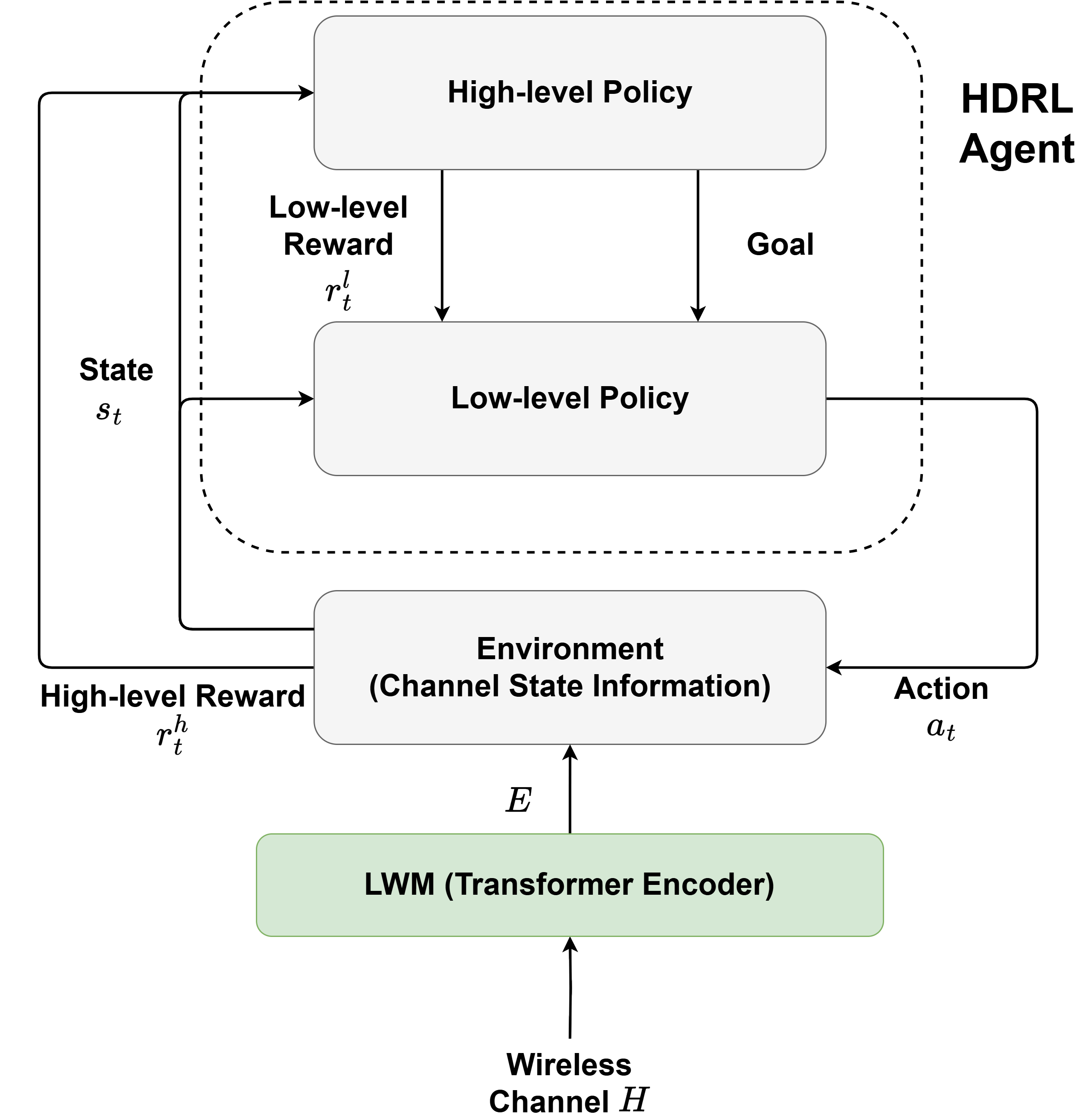}
    \caption{The proposed FM-HDRL framework architecture.}
    \label{fig:framework_architecture}
    \vspace{-4mm}
\end{figure}

We propose FM-HDRL to address the high-dimensional, non-convex optimization problem (P1). This framework leverages the ability of a fine-tuned LWM to extract compact channel features. This low-dimensional representation is then used by a two-level HDRL model for joint decision-making and optimization.
The overall architecture of the proposed framework is illustrated in \figurename{\ref{fig:framework_architecture}}.

\vspace{-1mm}
\subsection{FM for Channel Representation}
\vspace{-1mm}

LWM \cite{alikhani2024large} is a transformer-based foundation model pre-trained on large-scale channel datasets to learn universal representations of wireless channels. The LWM processes a raw channel matrix $\mathbf{H} \in \mathbb{C}^{X \times Y}$ by first flattening its real and imaginary parts into a single vector, which is then partitioned into a sequence of $P$ non-overlapping patches with a length of $L = \frac{2XY}{P}$.
Here, $ X $ and $ Y $ denote the dimensions of the channel matrix. 
Patches are linearly projected into a 128-dimensional latent space, generating embeddings fed into a transformer encoder. The model outputs a global classification (CLS) embedding representing the overall channel and 128-dimensional patch-level embeddings.

In this work, we fine-tune the final layer of the pre-trained LWM on our RIS-assisted channel dataset. The fine-tuned model generates a compact embedding vector $\mathbf{E} \in \mathbb{R}^{d_e}$ that encapsulates essential channel characteristics in a low-dimensional representation. This process enables the model to generate highly relevant, low-dimensional channel embeddings $\mathbf{E}$ for the HDRL model.

\vspace{-1mm}
\subsection{HDRL for Solving the Optimization Problem}
\begin{figure}
    \centering
    \includegraphics[width=0.9\linewidth]{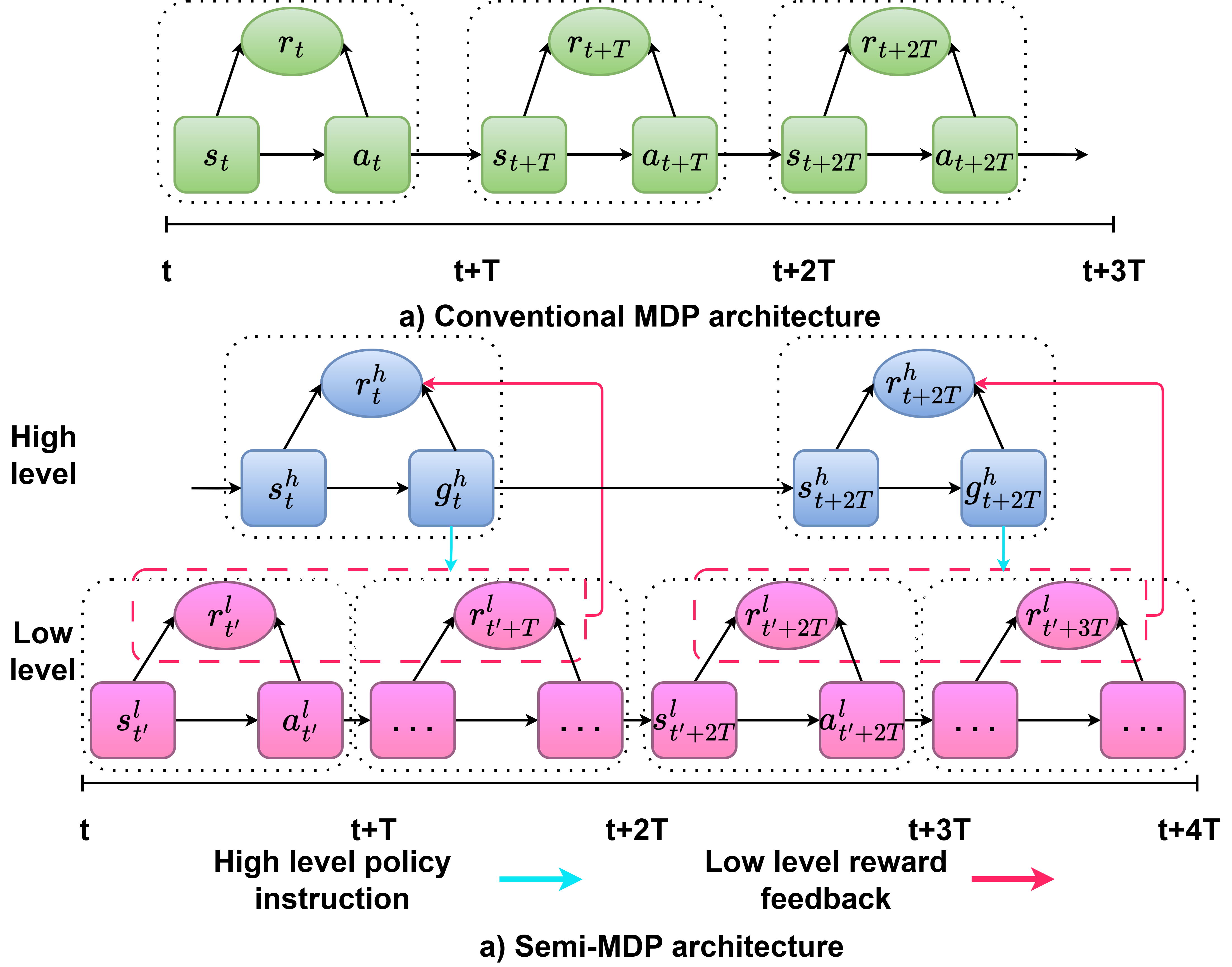}
    \caption{a) Conventional MDP architecture, and b) two-level SMDP architecture used in the HDRL framework.}
    \label{fig:smdp_model}
    \vspace{-4mm}
\end{figure}

\vspace{-1mm}
Solving the joint optimization problem using a DRL agent is often inefficient due to the high-dimensional action space. This approach, typically modeled as a Markov Decision Process (MDP), faces significant computational complexity and slow convergence. To overcome this, we adopt a hierarchical framework that improves both scalability and decision-making efficiency.
HDRL is modeled as a Semi-MDP (SMDP) and follows an off-policy learning strategy \cite{zhou2024cooperative}, as illustrated in \figurename{\ref{fig:smdp_model}}. The SMDP framework enables decision-making at different timescales through two hierarchical levels: a high-level meta-controller for long-term strategies and a low-level sub-controller for short-term actions.

\subsubsection{Meta-Controller (High-Level Policy)}
The meta-controller operates on a slower timescale, making strategic decisions every $T$ time slots. Its core function is to evaluate user locations and blockage status, and determine an appropriate strategy for the sub-controller. 

\vspace{-1mm}
\begin{itemize}

    \item \textbf{State ($s_t^h$):} At the beginning of each macro-slot, the meta-controller observes the environment state $s_t^h$, which includes the current locations of all $K$ users and the blockage conditions of the communication links. 

    \item \textbf{High-level goal ($g_t^h$):} Based on the observed state $s_t^h$, the meta-controller selects a high-level goal $g_t^h = \mathbf{b}_t = [b_{1,t}, b_{2,t}, \dots, b_{K,t}]$, where each binary variable $b_{k,t} \in \{0,1\}$ specifies the transmission mode for user $k$: $b_{k,t} = 1$ indicates that user $k$ should communicate via the RIS-assisted link, while $b_{k,t} = 0$ indicates the use of the RIS link. This goal serves as the strategic directive for the sub-controller during the macro-slot.

    \item \textbf{Reward ($R_t^h$):} The reward $R_{t}^{h}$ is the cumulative sum of rewards, $r_{t'}^{l}$, achieved by the sub-controller. It is designed to enforce the user fairness constraint (\ref{eq:fairness_constraint}) by incorporating a penalty, $P$, whenever the SE of any user drops below the minimum threshold, $R_{\text{min}}$, which formulated as $r_{t'}^{l} = \sum_{k=1}^{K} R_k(t') - P \cdot \mathbb{I}(\min_k R_k < R_{\text{min}})$, where $\mathbb{I}(\cdot)$ is the indicator function.
\end{itemize}

\subsubsection{Sub-Controller (Low-Level Policy)}
\vspace{-1mm}
Sub-Controller operates every time slot $t'$ within the macro-slot. Its task is to perform the optimization of the BS beamforming vectors $\mathbf{W}$ and the RIS phase-shift matrix $\mathbf{\Theta}$ parameters, all while adhering to the strategic goal set by the meta-controller.

\begin{itemize}
    \item \textbf{State ($s_{t'}^l$):} The state of the sub-controller is a combination of the channel embeddings $\mathbf{E}$ and the strategic goal from the meta-controller: $s_{t'}^l = \{\mathbf{E}, \mathbf{b}_t\}$. 

    \item \textbf{Action ($a_{t'}^l$):} The action is the selection of the BS beamforming vectors $\mathbf{W}$ and the RIS phase-shift matrix $\mathbf{\Theta}$. It learns to find the optimal configuration that maximizes the sum SE.

    \item \textbf{Reward ($r_{t'}^l$):} The sub-controller receives an immediate reward at each time step $t'$, which is the sum SE achieved in that slot, $r_{t'}^l = \sum_{k=1}^{K} R_k(t')$.
\end{itemize}

Algorithm \ref{alg:FMHDRL_final} outlines the overall training procedure.

\begin{algorithm}[htbp]
\caption{Training Procedure for the FM-HDRL Framework}
\label{alg:FMHDRL_final}
\SetAlgoNlRelativeSize{-1}
\KwIn{Initialize meta-actor, sub-actor, and corresponding critic networks.}
\KwIn{Initialize replay buffers for both levels of the hierarchy.}

\For{each episode}{
    \For{$t = 1, T+1, 2T+1, \dots$}{
        Observe initial user locations and blockage status $s_t^h$\;
        Meta-controller selects goal $g_t^h$ based on $s_t^h$\;
        Initialize cumulative reward $R_t^h = 0$\;
        
        \For{$t' = t$ to $t+T-1$}{
            Observe LWM channel embeddings $\mathbf{E}$\;
            Construct low-level state $s_{t'}^l = (\mathbf{E}, \mathbf{b}_t)$\;
            Sub-controller selects low-level action $a_{t'}^l = (\mathbf{W}, \mathbf{\Theta})$ based on $s_{t'}^l$\;
            Execute $a_{t'}^l$, observe immediate reward $r_{t'}^l$ and next low-level state $s_{t'+1}^l$\;
            Store sub-controller transition $(s_{t'}^l, a_{t'}^l, r_{t'}^l, s_{t'+1}^l)$ in its replay buffer\;
            Sample a minibatch and update sub-controller networks\;
        }
        
        Store meta-controller transition $(s_t^h, g_t^h, R_t^h, s_{t+T}^h)$ in its replay buffer\;
        Sample a minibatch and update meta-controller networks\;
    }
}

\end{algorithm}

%%%%%%%%%%%%%%%%%%%%%%%%%%%%%%%%%%%%%%%%%%%%%%%%%%%%%%%%%%%%

\vspace{-1mm}
\section{Simulation Results and Performance Comparison}
\vspace{-1mm}

In this section, we evaluate the performance of the proposed FM-HDRL framework. For performance evaluation, we benchmark our method against two baselines, namely FM-DRL and beam sweeping, with their details provided in Section V-C. 

\vspace{-1mm}
\subsection{Parameter Settings}
\vspace{-1mm}

The simulation environment consists of a single BS equipped with $N = 32$ antennas and an RIS with $M = 32$ passive reflecting elements, serving $K=50$ mobile users with $R_{\text{min}}=2$ bps/Hz. For each user position, channels are generated considering the 10 strongest propagation paths. 
We assume a frequency-flat channel model, wherein all subcarriers within an RB experience identical channel conditions during each transmission interval.

The LWM architecture follows the original model proposed in \cite{alikhani2024large}. 
It processes channel patches that are linearly embedded into a 128-dimensional feature space and augmented with positional encodings. The model consists of multiple Transformer encoder layers, each including multi-head self-attention, feed-forward networks, and residual normalization. A special CLS token is appended to obtain a global channel representation, while masked channel modeling is employed during pre-training to learn spatial and structural channel dependencies. We fine-tune the last layer of the pre-trained LWM on our RIS-specific scenario \cite{alkhateeb2019deepmimo}. The fine-tuning process uses a batch size of 64 and a learning rate of $10^{-5}$ with the AdamW optimizer. 
The current output of the LWM model is a tensor of shape [B, 128, 64], representing patch embeddings, with the last layer being a linear projection, where B refers to the batch size. During fine-tuning, this output is further projected to a lower dimension using another linear layer, and the loss function used is mean squared error (MSE).
The model processes the channel data with a patch size of $L=32$ and a total of $P=32$ patches per channel matrix. 

The deep deterministic policy gradient (DDPG)-based \cite{lillicrap2015continuous} agents for both the meta-controller and sub-controller are configured with specific learning parameters. The actor and critic networks for both levels use a learning rate of $10^{-4}$. The policy networks are composed of two hidden layers with 256 units each. We use a discount factor of $\gamma = 0.99$, with a replay buffer size of 500 for the meta-controller and 400 for the sub-controller.
The training is conducted for 100,000 episodes, with an update rate of 0.005. Furthermore, we define a smaller-scale time unit $t' = 0.5\,\text{ms} $, with each transmission interval $T = 10t'\,\text{ms}$.

\vspace{-1mm}
\subsection{DeepMIMO O1 Scenario}
\vspace{-1mm}

We utilize the "O1" outdoor urban scenario from the DeepMIMO dataset to ensure our evaluation reflects realistic deployment conditions \cite{alkhateeb2019deepmimo}.
This scenario is based on a realistic 3D ray-tracing model of a dense urban environment, providing a complex, accurate source of channel data. In this evaluation, "BS 15" is assigned as the BS, while "BS 7" serves as the RIS. 
The 'O1' scenario is particularly well-suited for our work as it includes over one million candidate user locations, covering a wide range of LoS and NLoS conditions.
We partition the dataset by allocating 70\% for fine-tuning, 15\% for validation, and the remaining 15\% for testing and performance evaluation.

\vspace{-1mm}
\subsection{Performance Analysis and Comparison}
\vspace{-1mm}

We evaluate the performance of the proposed FM-HDRL framework by assessing its training convergence behavior and communication performance, measured quantitatively by SE.
To assess its effectiveness, we compare our proposed approach against two baseline methods. The first is FM-DRL \cite{ghassemi2025foundation}, where a non-hierarchical DRL agent also utilizes $\mathbf{E}$ as input states. The second baseline is the beam sweeping approach, in which the BS and RIS sequentially scan a predefined codebook of beam configurations to identify the optimal setting. 
In our experimental setup, the beam sweeping baseline employs a codebook consisting of 1,024 beam pairs.

\subsubsection{Convergence Performance}
\vspace{-1mm}

\begin{figure}
    \centering
    \includegraphics[width=0.8\linewidth]{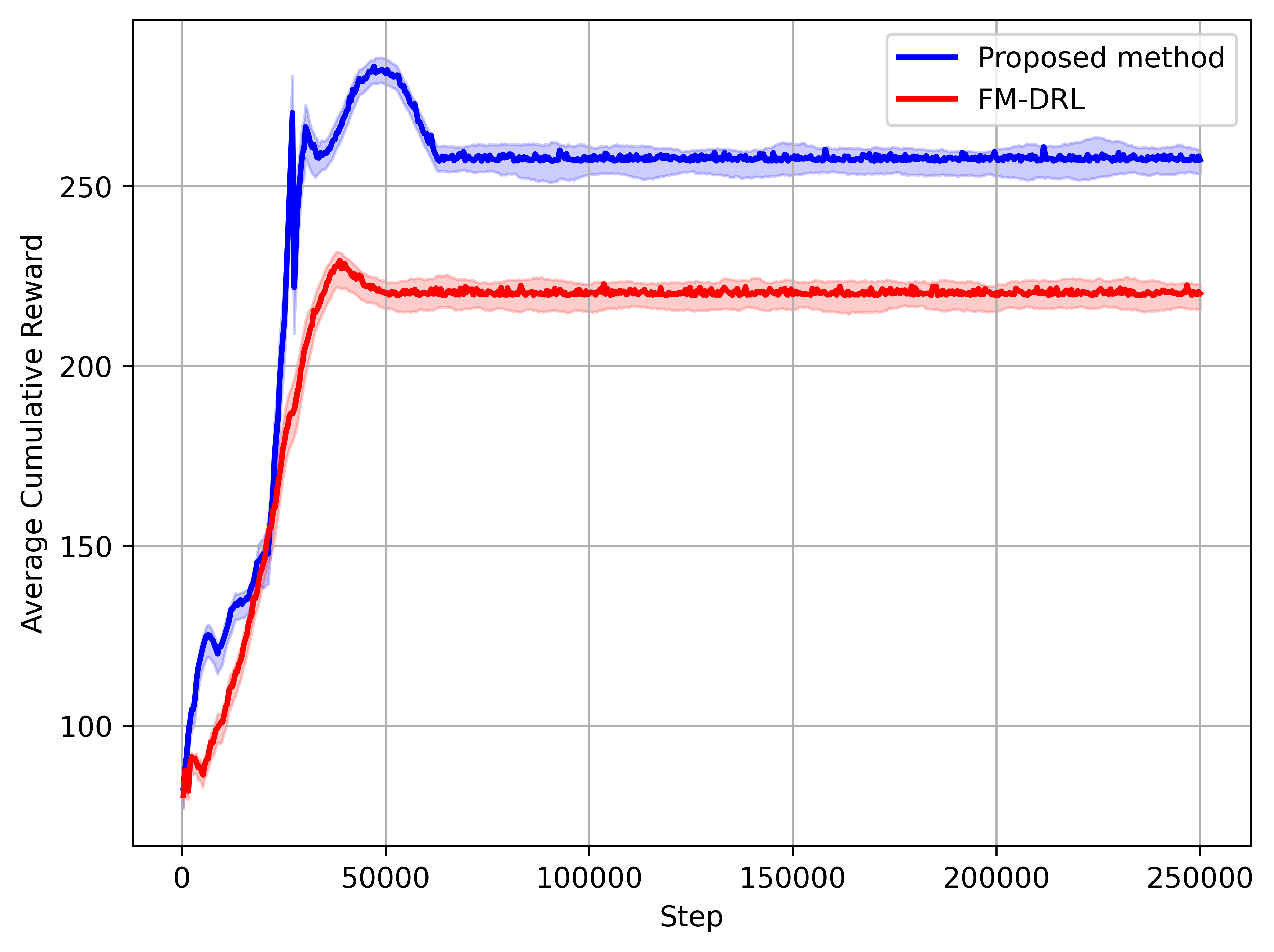}
    \caption{Average cumulative reward during training.}
    \label{fig:reward_convergence}
    \vspace{-4mm}
\end{figure}

\figurename~\ref{fig:reward_convergence} plots the average cumulative reward obtained by the proposed method and the FM-DRL agents. The proposed method achieves a higher cumulative reward and converges faster than its non-hierarchical counterpart. This performance gain stems from the hierarchical structure. The meta-controller handles high-level link selection, reducing decision complexity. Consequently, the sub-controller can focus exclusively on optimizing beamforming and RIS configuration within a simplified action space.
This faster convergence shows that the proposed hierarchical framework achieves high-performance policies with significantly lower sample complexity than the non-hierarchical counterpart.

\subsubsection{SE under Varying Transmit Power}
\begin{figure}
    \centering
    \includegraphics[width=0.8\linewidth]{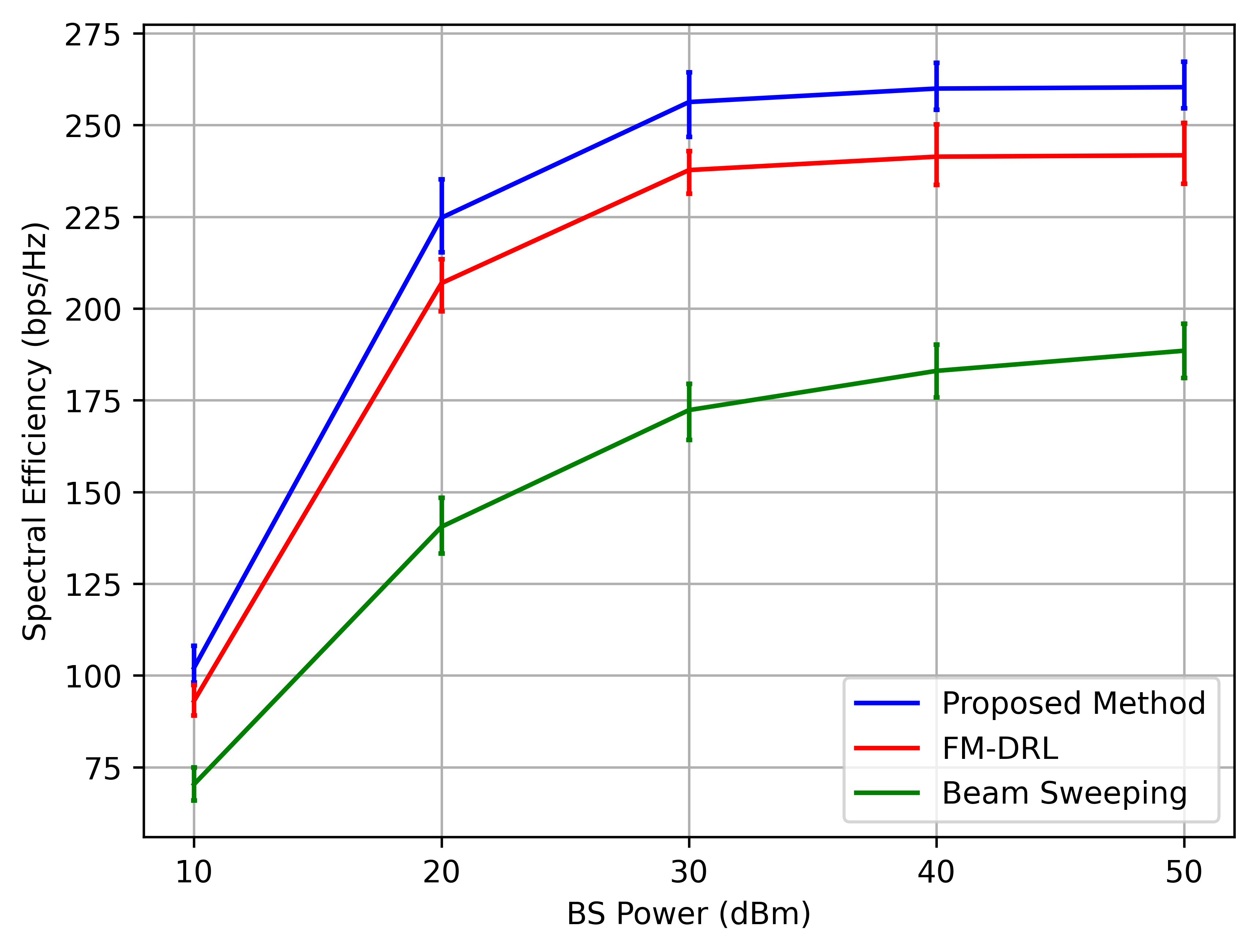}
    \caption{SE as a function of BS transmit power.}
    \vspace{-4mm}
    \label{fig:se_vs_power}
\end{figure}

\vspace{-1mm}
\figurename{\ref{fig:se_vs_power}} illustrates the SE of the three methods as a function of the BS transmit power. The proposed method consistently outperforms both baselines across all power levels. 
At a transmit power of 30 dBm, our method achieves an SE of 256.3 bps/Hz, which represents an improvement of approximately 7.82\% over the 237.7 bps/Hz achieved by FM-DRL and a significant gain of around 48.66\% compared to the 172.4 bps/Hz achieved through beam sweeping.
This demonstrates that the proposed method is more effective at jointly optimizing the beamforming and phase shifts to leverage the available power budget. The limited performance of the beam sweeping method arises from two key factors: the practical constraint of using quantized codebooks and the lack of channel-aware adaptation, which prevents the method from responding effectively to real-world channel dynamics.

\subsubsection{Scalability with RIS Size}
\begin{figure}
    \centering
    \includegraphics[width=0.8\linewidth]{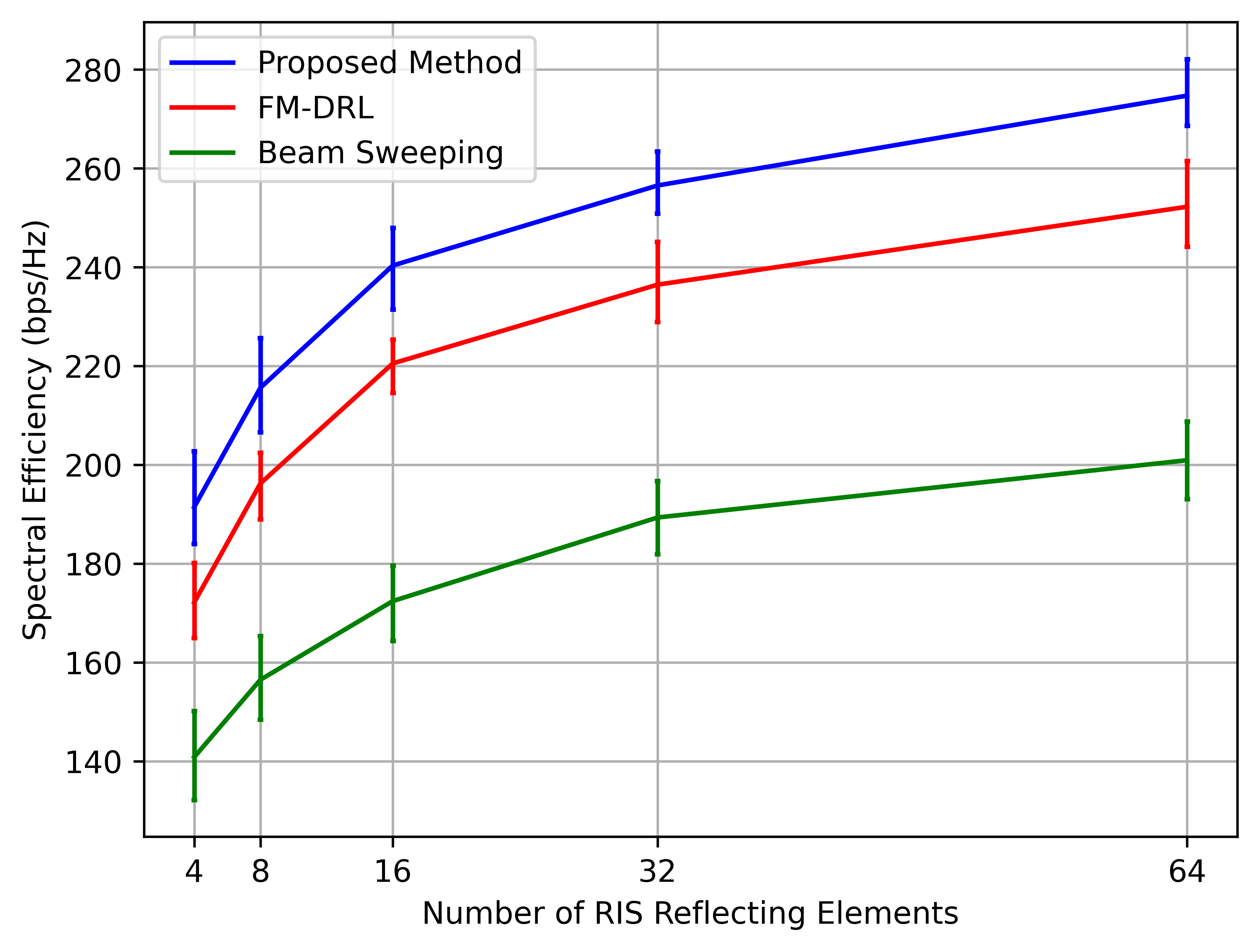}
    \caption{SE as a function of the number of RIS reflecting elements.}
    \vspace{-4mm}
    \label{fig:se_vs_ris}
    
\end{figure}

\vspace{-1mm}
To evaluate the scalability of our approach, we measure the SE while increasing the number of reflecting elements in the RIS at a transmit power of 30 dBm.  
As shown in \figurename~{\ref{fig:se_vs_ris}}, the performance of all methods improves with a larger RIS, but the gains achieved by the proposed method are more prominent. With 64 RIS elements, our framework achieves an SE of 274.7 bps/Hz, a 9\% improvement over the FM-DRL. Notably, the performance gap between the proposed method and beam sweeping widens as $M$ increases. The beam sweeping method, which relies on a fixed codebook, struggles to cope with the exponentially growing search space. 

\subsubsection{Computational Complexity Analysis}
\vspace{-1mm}

The computational cost of FM-HDRL arises from LWM inference and HDRL training. HDRL substantially reduces action-space complexity compared to DRL. During deployment, the meta-controller is invoked every $T$ slots, while the sub-controller leverages $\mathbf{E}$. Compared to iterative optimization and exhaustive beam sweeping, FM-HDRL offers a favorable performance--complexity trade-off for centralized RAN controllers.

\subsubsection{Limitations and Future Work}
\vspace{-1mm}

The evaluation is limited to a single urban scenario and assumes perfect user location and blockage information. While results indicate the effectiveness of FM-HDRL for RIS-assisted networks, future work will examine robustness to imperfect CSI and localization errors, latency to assess deployment in real RAN controllers, diverse deployment scenarios, hardware impairments, and scalability to larger networks.

%%%%%%%%%%%%%%%%%%%%%%%%%%%%%%%%%%%%%%%%%%%%%%%%%%%%%%%%%%%%

\vspace{-1mm}
\section{Conclusion}
\vspace{-1mm}

In this paper, we introduced FM-HDRL to address the complex problem of joint beamforming and phase shift control in blockage-aware, RIS-assisted wireless networks. By fine-tuning the LWM to generate concise and robust channel representations and integrating it with a two-level HDRL agent, the proposed approach effectively addresses the challenges posed by the high-dimensional and non-convex nature of the optimization problem. Our simulations confirm that the proposed method converges faster, achieves higher SE, and offers superior scalability compared to both non-hierarchical DRL and beam-sweeping baselines. These findings highlight the potential of integrating large-scale models with HDRL to unlock the full capabilities of next-generation wireless systems.

\vspace{-1mm}
\section*{Acknowledgment}
\vspace{-1mm}
This work has been supported by NSERC Canada Research Chairs program, MITACS, and Ericsson. We also wish to honor the memory of our co-author, Han, whose invaluable contributions and dedication in writing this paper were deeply appreciated. Her efforts and spirit will always be remembered.

\vspace{-1mm}
\bibliographystyle{IEEEtran}
\bibliography{bibliography.bib}

\end{document}